## First-principles calculations of the structural, elastic and electronic properties of $MN_xC_{1-x}$ ($M$=Ti, Zr, Hf; $0 \le x \le 1$) carbonitrides at ambient and elevated hydrostatic pressure

**V. Krasnenko, M.G. Brik***

Institute of Physics, University of Tartu, Riia 142, Tartu 51014, Estonia

**Abstract**

The structural, electronic, and elastic properties of three mixed transition metal carbonitrides $TiN_xC_{1-x}$, $ZrN_xC_{1-x}$, and $HfN_xC_{1-x}$ ($0 \le x \le 1$) with the rock-salt structure were calculated at ambient and elevated up to 50 GPa hydrostatic pressures in the framework of the density functional theory methods. The lattice constants, densities, and bulk moduli of the considered compounds were shown to behave as linear functions of the nitrogen concentration $x$. The obtained linear dependencies of all these parameters allow for getting their estimates at any value of $x$ in the range from 0 to 1. Gradual enhancement of the ionicity of the chemical bonds with gradual replacement of carbon by nitrogen was demonstrated by calculating the bond orders and electron density difference distributions.

## 1. Introduction

Many transition metal (TM) carbides and nitrides exhibit a number of unique physical properties, e.g. chemical stability, high melting temperature, and corrosion resistance [1, 2, 3, 4, 5,6]. It is worthwhile specially noting that the hardness of these materials is comparable to that one of diamond [7]. Such an attractive combination of remarkably good characteristics makes them the materials of choice in many important industrial and technological applications, like coating of rotating and cutting tools, especially those operating at extremely high pressures and temperatures. Many of the TM carbides and nitrides crystallize in the rock salt structure that is typical for compounds with ionic bond. Additionally, TM carbides and nitrides possess good electrical and thermal conductivity that is characteristic of compounds with metallic bond [8].

* Corresponding author. Tel.: +372 7374751; fax: +372 738 3033. E-mail address: brik@fi.tartu.ee (M.G. Brik).

In some cases, the TM carbonitrides (ternary stoichiometric compounds with general chemical formula $MN_xC_{1-x}$, $M$=3d, 4d, or 5d TM or ternary nonstoichiometric compounds $MN_xC_y$, sometimes with vacancies of any of these chemical elements) are more preferable than their corresponding pure carbides or nitrides. For example, the coatings of chromium carbonitride are becoming more and more interesting for wear protection applications due to their increased hardness and improved wear performance compared to chromium nitride hard coatings [9]. Also, the hardness enhancement of titanium carbonitrides had been found, when simultaneously combining Ti with C and N [10, 11, 12]. An electronic mechanism of such a hardness enhancement was proposed based on the *ab initio* pseudopotential calculations [13].

Alloying in TM carbides and nitrides has not been much studied experimentally. Very few studies have been performed on the phase equilibrium in the HfCN system [14]. The electronic properties of the titanium carbonitride alloys were calculated by Zhukov *et al* [15], and the *ab initio* studies of the structural stability and elastic stiffness of these alloys can be found in Ref. [16]. If the ternary TM carbonitrides are considered, the overall picture of their physical properties, including chemical bonding peculiarities, becomes very complicated, resulting in difficulties in interpretation of their electronic structure [17]. Many of the TM carbides and nitrides are capable of forming unlimited solid solutions with each other [8], whose physical and chemical properties depend to a large extent on their composition and/or presence of vacancies. Some properties of the carbonitrides, as a function of the C/N ratio, are non-monotonous and can be understood by considering their electronic structure in details [17].

In the present paper we continue our previous studies of the TM mononitrides [18] and monocarbides [19] by presenting the results of a consistent analysis of the structural, elastic and electronic properties of the $TiN_xC_{1-x}$, $ZrN_xC_{1-x}$, and $HfN_xC_{1-x}$ ternary compounds as they are modified by a gradual replacement of the carbon atoms by nitrogen in the whole range of possible concentrations, i.e. going from pure carbides to pure nitrides. All calculated properties were obtained not only at the ambient, but at the elevated pressures as well, which – if taken together with the complete study of the concentration effects – helps get complementary picture of a wide range of the physical properties of these materials. The paper is composed as follows: in the next section the structures of the pure and mixed compounds studied in the present work are described along with the main calculating settings. After having finished the propedeutic part of the paper, we proceed with the description and discussion of all obtained results before we reach a short summary of the main results.

## 2. Crystal structures and computational details

The neat TiC, TiN, ZrC, ZrN, HfC and HfN all crystallize in the Fm-3m space group (No. 225) with four formula units in one unit cell. The experimental lattice constants of these pure carbides and nitrides are collected in Table 1, whereas Fig. 1 displays one unit cell of these materials with four TM atoms and four carbon (nitrogen) atoms.

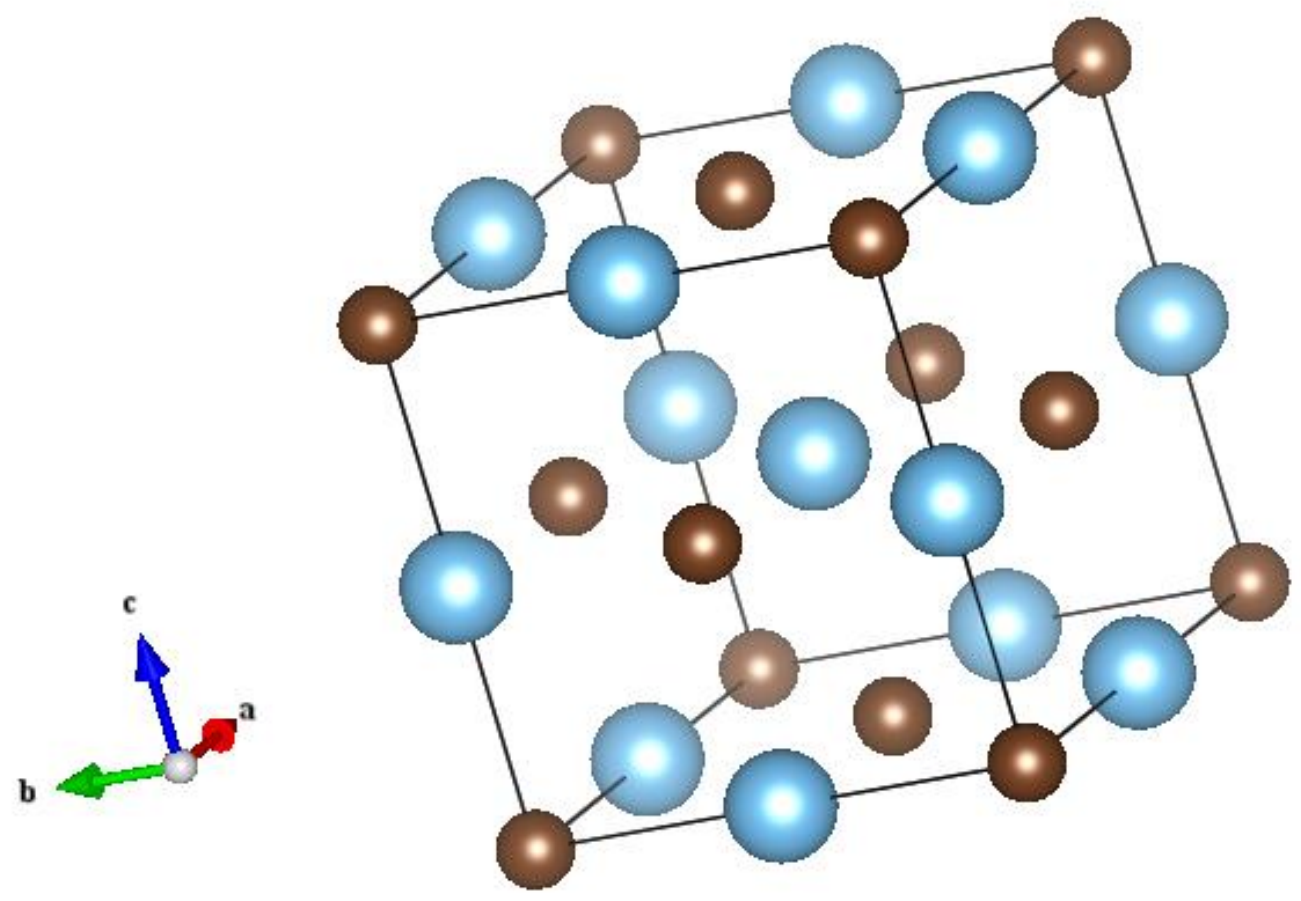

Fig. 1. One unit cell of TiC. The Ti and C atom are shown by bigger and smaller spheres, correspondingly. Drawn with VESTA [20].

The structures of other nitrides and carbides considered in the present paper are similar and thus are not shown. The carbonitrides $MN_xC_{1-x}$ ($M$=Ti, Zr, Hf, x=0, 0.25, 0.50, 0.75, 1.0) were simulated starting with one unit cell of a pure corresponding carbide gradually substituting the carbon atoms with nitrogen, which would correspond to the $M_4C_4$, $M_4NC_3$, $M_4N_2C_2$, $M_4N_3C$, $M_4N_4$ clusters.

All presented calculations were performed using the CASTEP module [21] of Materials Studio in two independent computational runs, either in the generalized gradient approximation (GGA) with the Perdew-Burke-Ernzerhof functional [22] or in the local density approximation (LDA) with the Ceperley-Alder-Perdew-Zunger (CA-PZ) functional [23, 24]. The plane-wave basis energy cutoff was chosen to be 310 eV for $TiN_xC_{1-x}$ and 280 eV for $ZrN_xC_{1-x}$ and $HfN_xC_{1-x}$. The Monkhorst-Pack scheme *k*-points grid sampling was set at 10×10×10 for the Brillouin zone. The convergence parameters were set as follows: total energy tolerance $-10^{-5}$ eV/atom, maximum force tolerance 0.03 eV/nm, maximum stress 0.05 GPa, and maximum displacement 0.001 Å. The electronic configurations for all involved chemical elements were as follows: $2s^2 2p^2$ for carbon, $2s^2 2p^3$ for nitrogen, $3s^2 3p^6 3d^2 4s^2$ for titanium, $4s^2 4p^6 4d^2 5s^2$ for zirconium and $5d^2 6s^2$ for hafnium.

## 3. Results and discussions

### 3.1. Structural properties

The first necessary step of any first-principles calculations is getting the optimized crystal structures closely related to the existing (if any) reliably determined experimental crystallographic parameters. Good agreement between the theoretical and empirical structural data (with a typical discrepancy of not more than a few percent at most) allows for performing the next round of calculations pertaining to the electronic and elastic properties starting with the optimized structures. To follow this route, the structural optimization was done for all considered pure carbides and nitrides and mixed ternary compounds (fifteen systems in total: $M$C, $M$N$_{0.25}$C$_{0.75}$, $M$N$_{0.50}$C$_{0.50}$, $M$N$_{0.75}$C$_{0.25}$, $M$N, $M$=Ti, Zr, Hf).

The calculated and experimental structural properties are collected in Table 1. For comparison, we have included the results of other *ab initio* calculations found in the literature when available. It can be seen from the Table that the lattice constants decrease smoothly with a similar step when moving from carbides to nitrides.

Table 1. Lattice constants $a$ (in Å) for three groups of the $TiN_xC_{1-x}$, $ZrN_xC_{1-x}$ and $HfN_xC_{1-x}$ carbonitrides. Linear dependencies of the lattice constants on the nitrogen concentration $x$ ($0 \le x \le 1$) are also given.

| Compounds | Exp. | Calc. (this work) | | Calc. (other data) |
|---|---|---|---|---|
| | | LDA | GGA | |
| TiC | 4.327 [a] | 4.27444 | 4.34906 | 4.319 [b], 4.332 [c], 4.272 [d] |
| $TiC_{0.75}N_{0.25}$ | | 4.24799 | 4.32160 | 4.295 [b], 4.25 [d] |
| $TiC_{0.50}N_{0.50}$ | | 4.22409 | 4.29824 | 4.275 [b], 4.22 [d] |
| $TiC_{0.25}N_{0.75}$ | | 4.20156 | 4.27672 | 4.254 [b], 4.21 [d] |
| TiN | 4.235 [a] | 4.18464 | 4.25049 | 4.237 [b], 4.176 [d] |
| | | 4.27176 (1-$x$)+4.18132 $x$ | 4.34763 (1-$x$)+4.25082 $x$ | |
| ZrC | 4.692 [e] | 4.647616 | 4.719675 | 4.71 [f] |
| $ZrC_{0.75}N_{0.25}$ | 4.67 [f] | 4.62515 | 4.698762 | 4.68 [f] |
| $ZrC_{0.50}N_{0.50}$ | 4.65 [f] | 4.602659 | 4.677556 | 4.66 [f] |
| $ZrC_{0.25}N_{0.75}$ | 4.63 [f] | 4.572804 | 4.651234 | 4.64 [f] |
| ZrN | 4.575 [e] | 4.559402 | 4.639222 | 4.66 [f] |
| | | 4.64728 (1-$x$)+ 4.55577 $x$ | 4.71898(1-$x$)+4.63561 $x$ | |
| HfC | 4.64 [g] | 4.64250 | 4.71487 | 4.657 [d], 4.7076 [h] |
| $HfC_{0.75}N_{0.25}$ | | 4.62680 | 4.69782 | 4.619 [d], 4.691 [h] |
| $HfC_{0.50}N_{0.50}$ | | 4.61006 | 4.68143 | 4.592 [d], 4.676 [h] |
| $HfC_{0.25}N_{0.75}$ | | 4.59343 | 4.66493 | 4.568 [d], 4.658 [h] |
| HfN | 4.52 [i] | 4.57777 | 4.65051 | 4.537 [d], 4.6455 [h] |
| | | 4.64265 (1-$x$)+4.57761 $x$ | 4.71426 (1-$x$)+4.64952 $x$ | |

[a] Ref. [25]
[b] Ref. [12]
[c] Ref. [26]
[d] Ref. [27]
[e] Ref. [28]
[f] Ref. [29]
[g] Ref. [30]
[h] Ref. [31]
[i] Ref. [32]

A visualization of the existing trend in the calculated structural properties of the considered ternary carbonitrides is offered by Fig. 2, which exhibits the perfectly linear behavior of the calculated lattice constants as the functions of the ternary alloys composition, as can be anticipated according to the Vegard's law. The linear fits of these results are presented both in Table 1 and Fig. 2; $x$ is a non-dimensional concentration, and the obtained result is the lattice constant in Å.

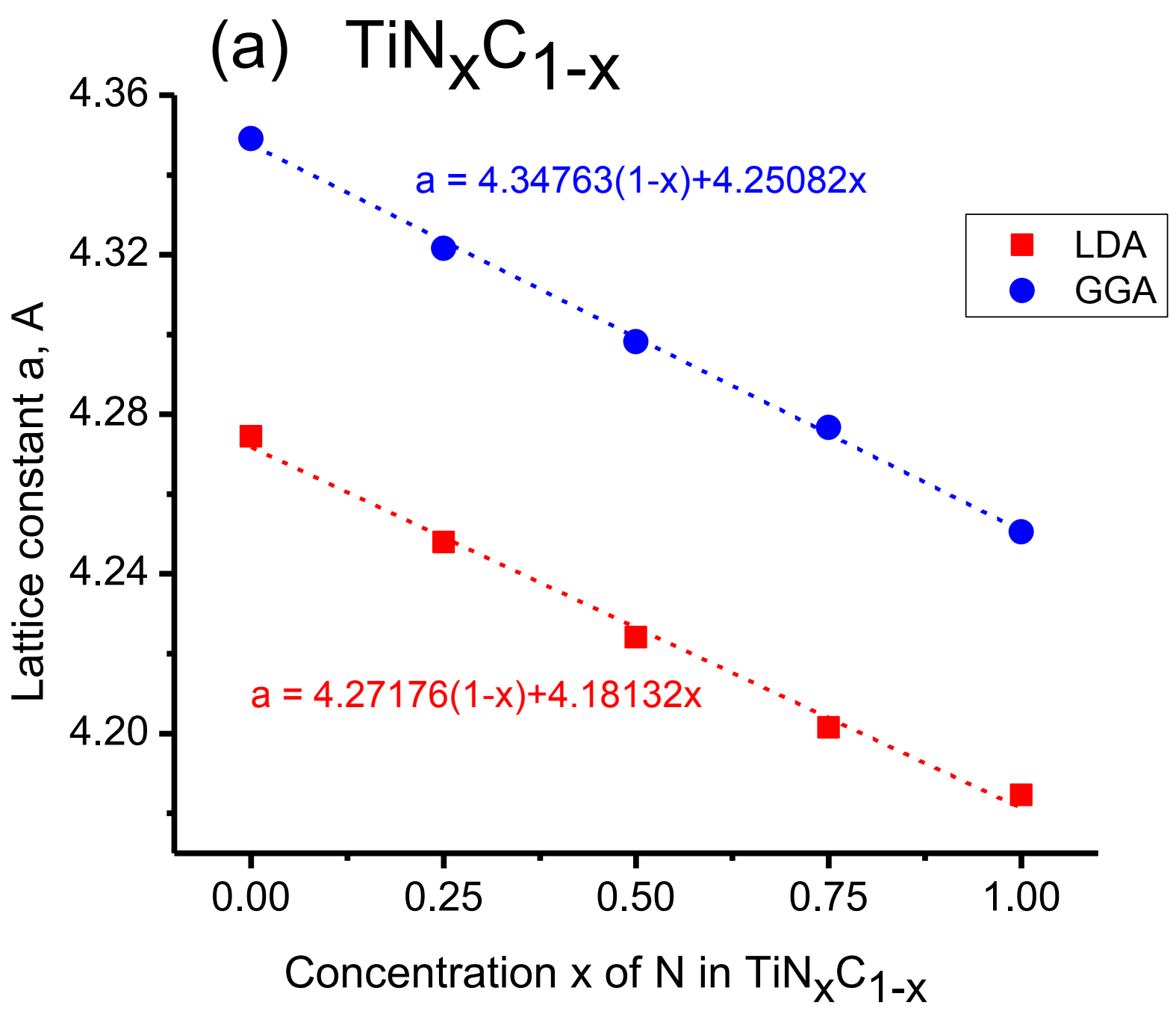

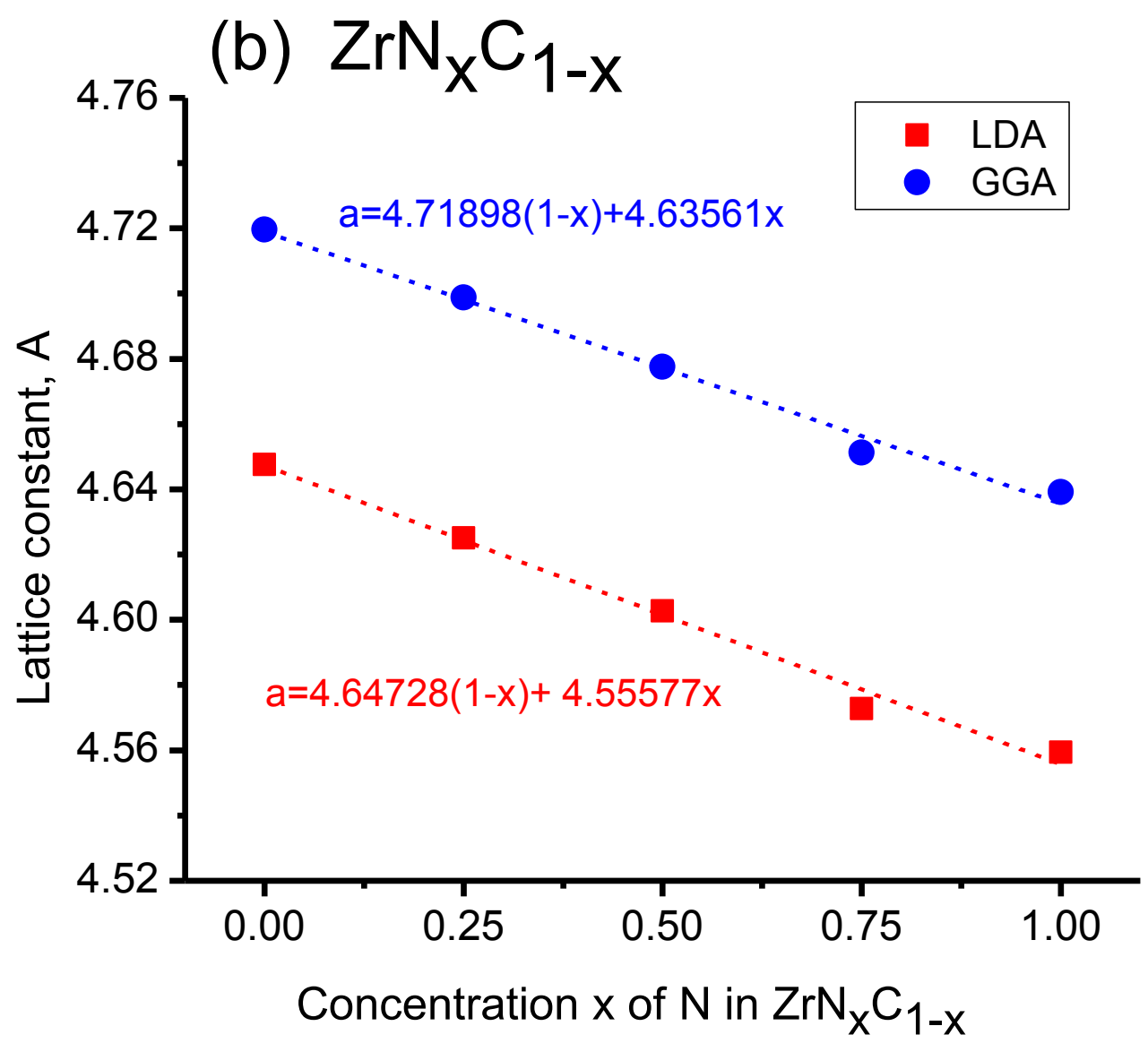

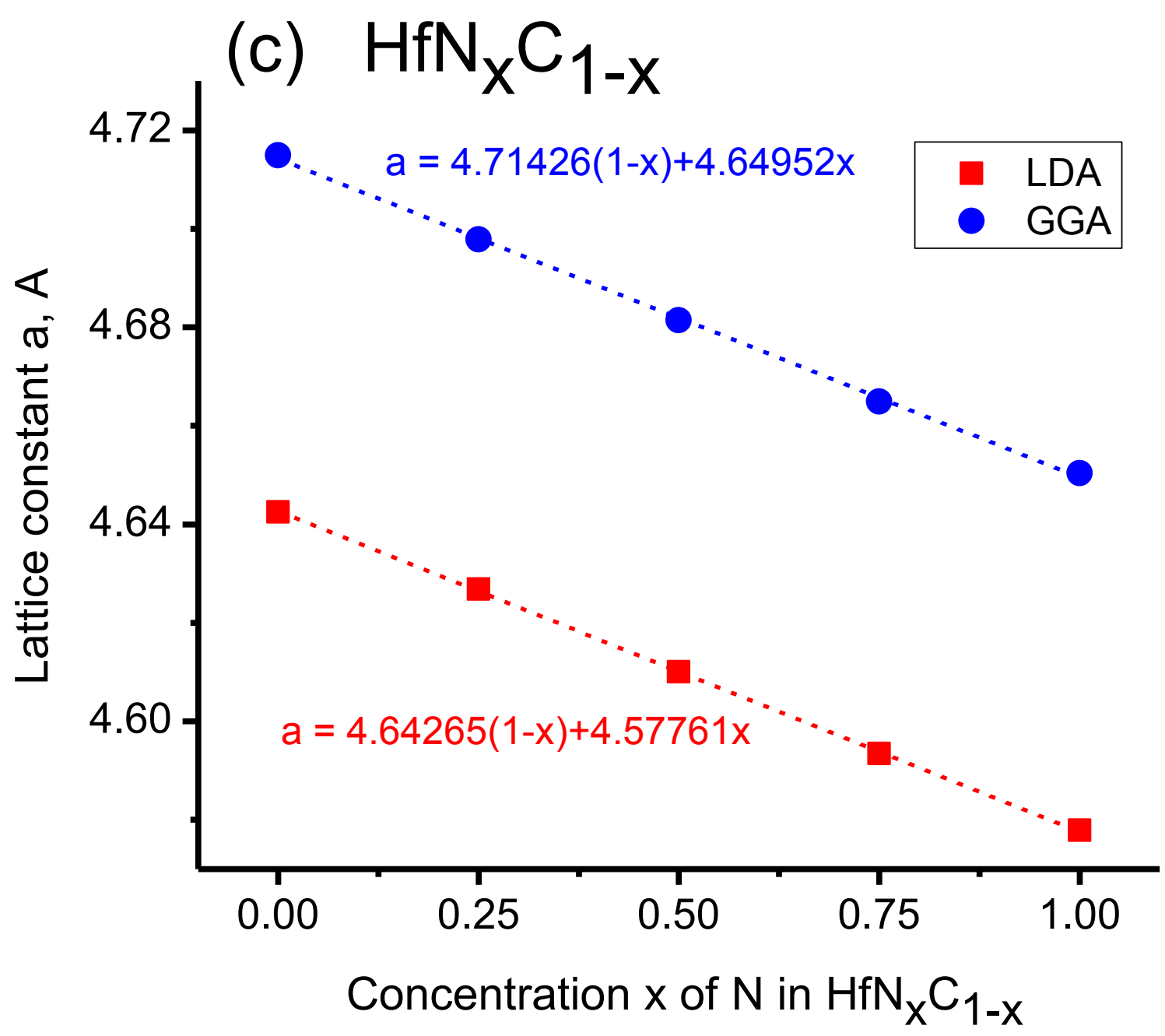

Fig. 2. Dependence of the calculated lattice constants on the nitrogen composition $x$ for $TiN_xC_{1-x}$ (a), $ZrN_xC_{1-x}$ (b) and $HfN_xC_{1-x}$ (c) carbonitrides.

The density of all carbonitrides was also calculated; the results are presented in Fig. 3. The density excellently follow the linear trend across the whole range of the nitrogen concentration.

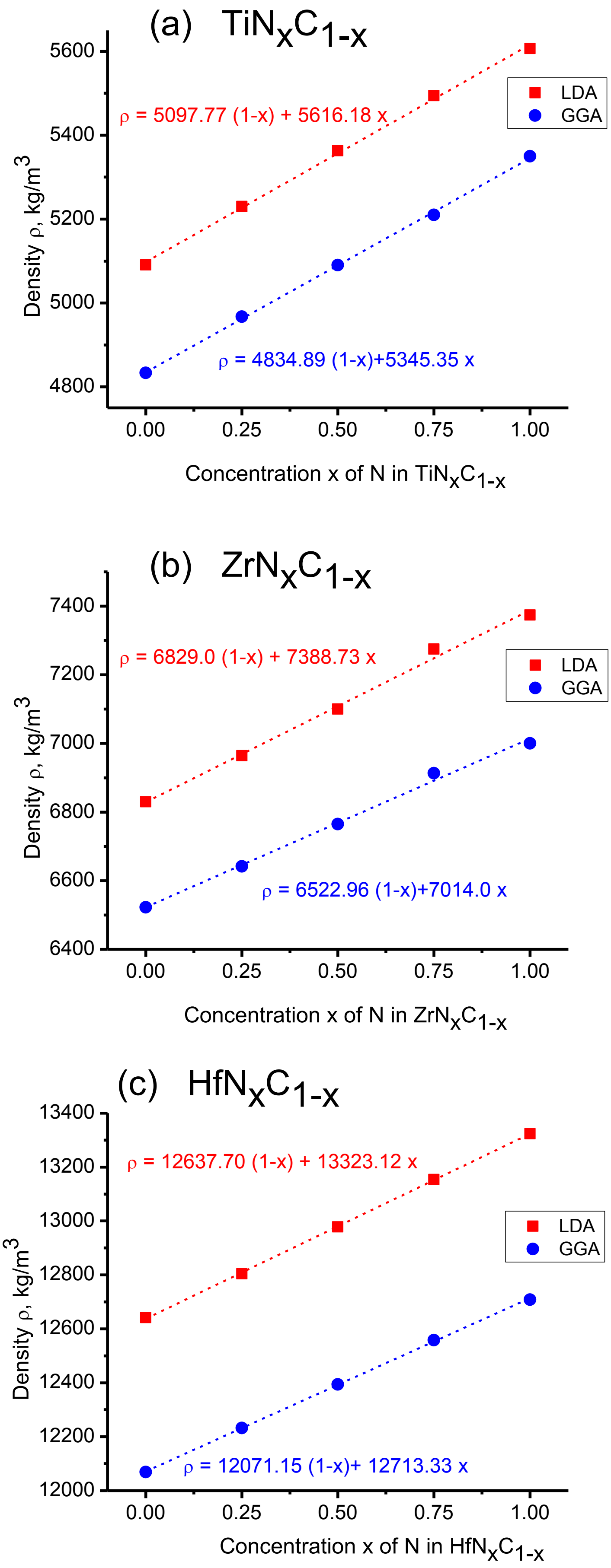
(a) TiNxC1-x
ρ = 5097.77 (1-x) + 5616.18 x
ρ = 4834.89 (1-x)+5345.35 x
LDA
GGA
Density ρ, kg/m3
Concentration x of N in TiNxC1-x
(b) ZrNxC1-x
ρ = 6829.0 (1-x) + 7388.73 x
ρ = 6522.96 (1-x)+7014.0 x
LDA
GGA
Density ρ, kg/m3
Concentration x of N in ZrNxC1-x
(c) HfNxC1-x
ρ = 12637.70 (1-x) + 13323.12 x
ρ = 12071.15 (1-x)+ 12713.33 x
LDA
GGA
Density ρ, kg/m3
Concentration x of N in HfNxC1-x

Fig. 3. Dependence of the calculated density $\rho$ on the nitrogen composition $x$ for $TiN_xC_{1-x}$ (a), $ZrN_xC_{1-x}$ (b) and $HfN_xC_{1-x}$ (c) carbonitrides.

The fits equations are shown in the $a_{carb}(1-x) + a_{nitr}x$ form, $0\leq x \leq 1$, with the $a_{carb}$ and $a_{nitr}$ being the coefficients (measured in Å) corresponding to the lattice constants of the pure carbide and pure nitride. Such representation sets clearly the upper and lower limits for the lattice constants variation and separates the "contributions" of pure nitrides and carbides into the properties of a mixed compound.

3.2. Elastic properties

Calculations of the elastic properties for all studied materials were performed for the optimized crystal structures. The calculated bulk moduli are all shown in Table 2. The first conclusion, which can be drawn from these data, is that the nitrides are harder than carbides, and when the carbon is gradually substituted by nitrogen, the values of the bulk moduli are increasing. The character of these changes becomes clear after inspection of Fig.4: the trend is linear again, in accordance with the Vegard's law. The quality of linear fits ($x$ is a dimensionless concentration $0\leq x \leq 1$; $B$ is the bulk modulus in GPa) is good, which ensures a possibility of estimating the bulk modulus value for any nitrogen concentration $x$ in the range from 0 to 1. We also note that the calculated bulk moduli are highly consistent with other available data (Table 2).

The greater values of $B$ for nitrides in comparison with those for carbides can be related to a greater number of valence electrons (9 in nitrides and 8 in carbides), which agrees with the previous literature data [19, 27]. This statement can be also applied to the ternary compounds with mixed composition of N and C as well.

For a cubic crystal, one needs to know the values of three elastic constants $C_{11}$, $C_{12}$, $C_{44}$, in order to calculate the response of such a material to external deformations. Direct calculations of the elestic constants tensor components are avialable in CASTEP; the obtained results are presented in Fig. 5 and Table 3. The variation of $C_{ij}$ with the nitrogen concentration $x$ is not as smooth as those of the lattice constants or bulk moduli, although the overall trend can be formulated as follows: in the entire range of concentration $x$ the values of the $C_{11}$ and $C_{12}$ constants grow up with increasing $x$, whereas the $C_{44}$ constant shows an opposite behavior. Such a non-monotonic variation of $C_{44}$ was also noticed earlier [27, 31]. The $C_{11}$ constant describes elastic response to the axial deformation (along the $a$, $b$, $c$ crystallographic axes).

Since all materials considered in the present paper are cubic and the chemical bonds are parallel to these axes (Fig. 1), increased values of $C_{11}$ are directly related to the shortened chemical bonds, which become more rigid under compression.

Table 2. Bulk moduli $B$ (in GPa) for three groups of the $TiN_xC_{1-x}$, $ZrN_xC_{1-x}$ and $HfN_xC_{1-x}$ carbonitrides. Linear dependencies of the bulk moduli on the nitrogen concentration $x$ ($0 \leq x \leq 1$) are also given.

| Compound | Calculated (this work) | | Exper. | Other calc. |
|---|---|---|---|---|
| | LDA | GGA | | |
| TiC | 258.21 | 225.28 | 240 [a] | 250 [b], 267 [c], 277 [d], 281 [e], 220 [f], 278 [g] |
| $TiC_{0.75}N_{0.25}$ | 270.52 | 237.35 | | 257.8 [b] |
| $TiC_{0.50}N_{0.50}$ | 280.89 | 245.94 | | 265.1 [b] |
| $TiC_{0.25}N_{0.75}$ | 295.15 | 253.76 | | 271.4 [b] |
| TiN | 302.97 | 256.69 | 288 [h] | 277 [b] |
| | 258.85 $(1-x)$+304.0 $x$ | 227.96 $(1-x)$+259.65 $x$ | | |
| ZrC | 234.00 | 208.09 | | 219 [i], 232 [i], 247 [i] |
| $ZrC_{0.75}N_{0.25}$ | 241.30 | 211.42 | | 227 [i] |
| $ZrC_{0.50}N_{0.50}$ | 248.72 | 212.92 | | 233 [i] |
| $ZrC_{0.25}N_{0.75}$ | 254.69 | 218.01 | | 239 [i] |
| ZrN | 268.16 | 226.14 | | 245 [i], 246 [i], 285 [i] |
| | 233.03 $(1-x)$+265.71 $x$ | 206.78 $(1-x)$+223.86 $x$ | | |
| HfC | 262.52 | 237.70 | | 233 [b], 284 [j] |
| $HfC_{0.75}N_{0.25}$ | 277.56 | 247.82 | | 255.7 [b], 273.6 [j] |
| $HfC_{0.50}N_{0.50}$ | 292.37 | 254.73 | | 262.6 [b], 275.4 [j] |
| $HfC_{0.25}N_{0.75}$ | 297.01 | 259.68 | | 267.6 [b], 273.6 [k] |
| HfN | 305.27 | 267.87 | 306 [i] | 272.8 [b], 283.4 [l] |
| | 264.92 $(1-x)$+311.05 $x$ | 239.31 $(1-x)$+267.44 $x$ | | |

[a] Ref. [33]
[b] Ref. [31]
[c] Ref. [34]
[d] Ref. [35]
[e] Ref. [27]
[f] Ref. [36]
[g] Ref. [37]
[h] Ref. [38]
[i] Ref. [29]
[k] Ref. [39]
[l] Ref. [27]

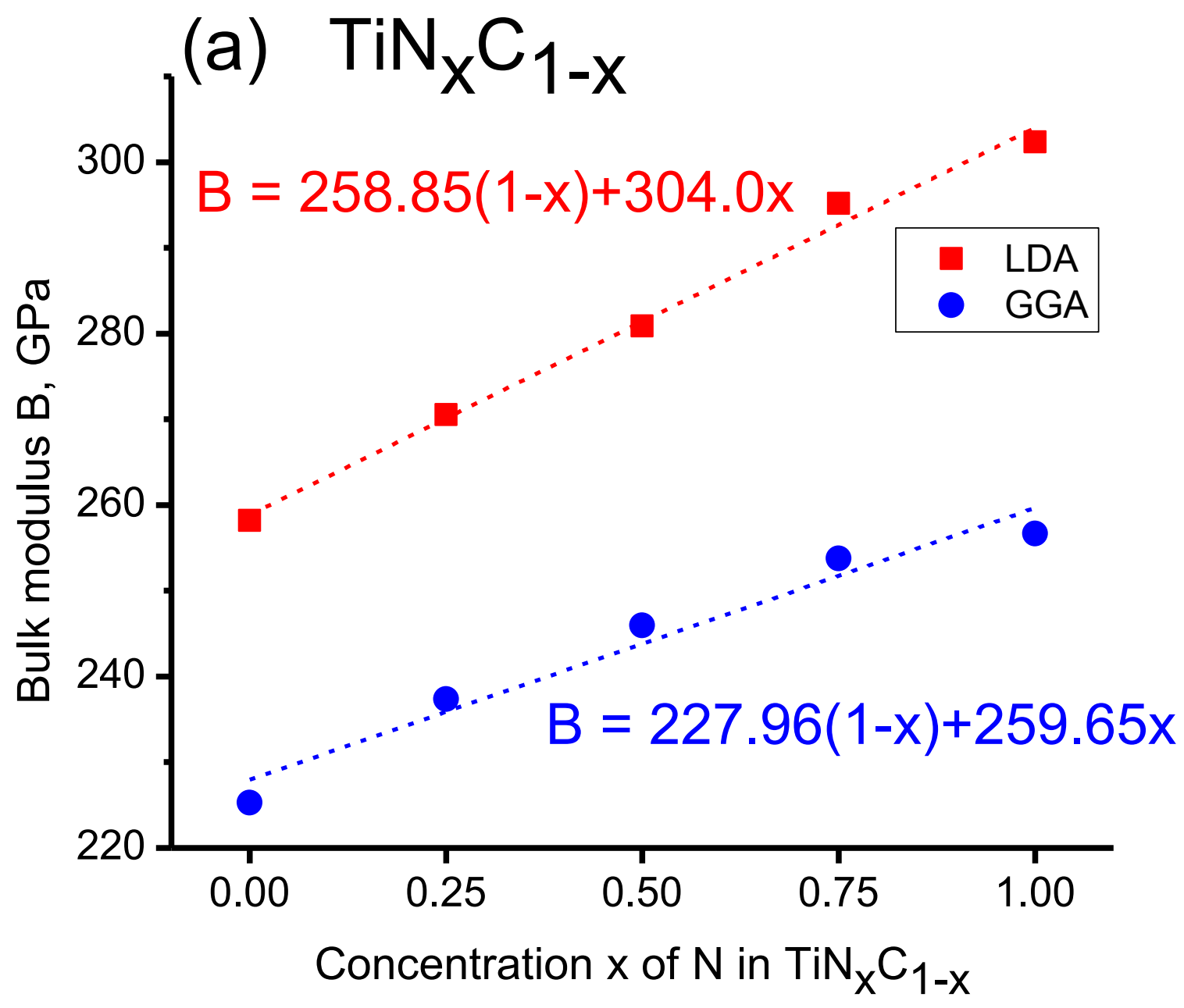

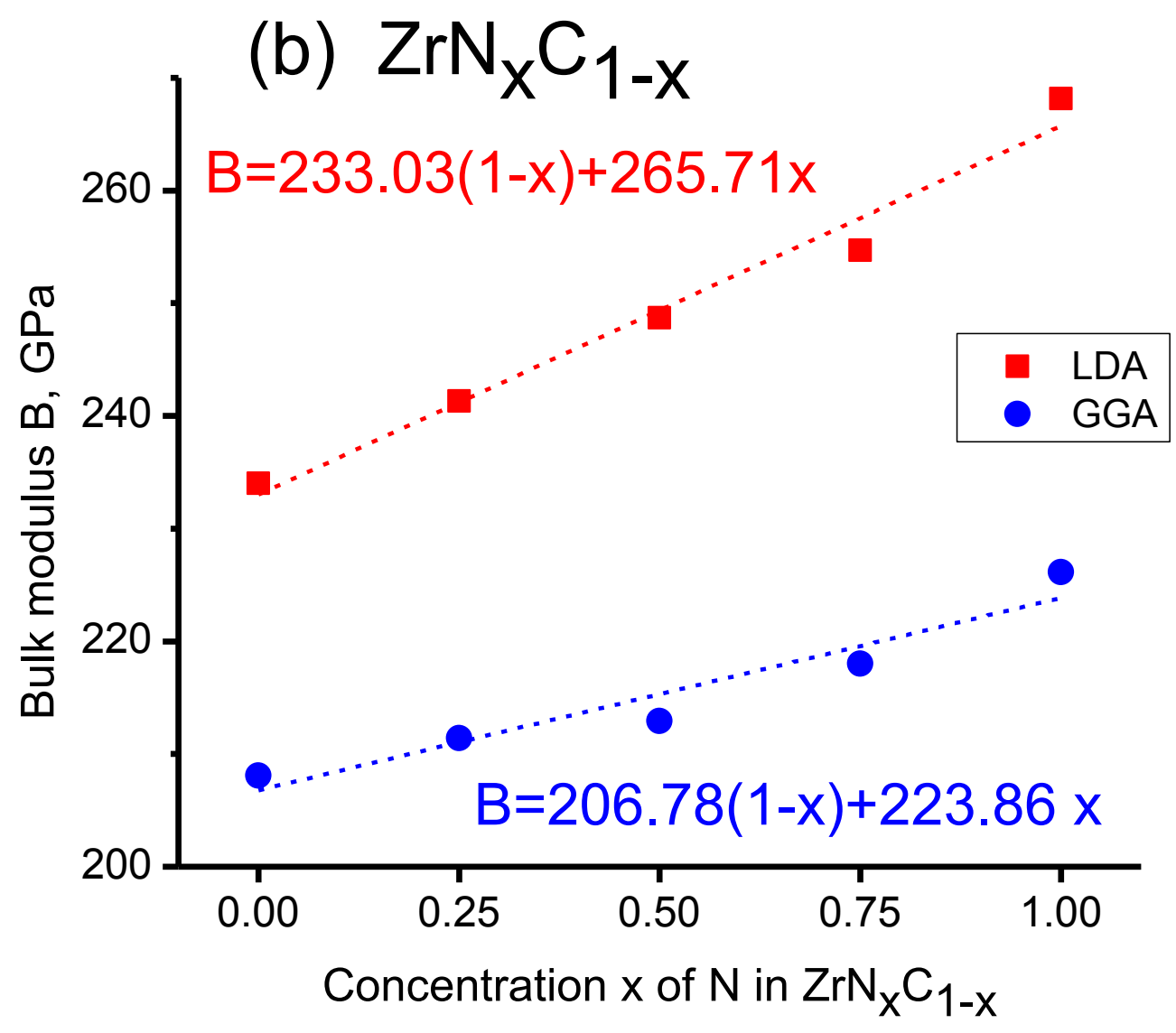

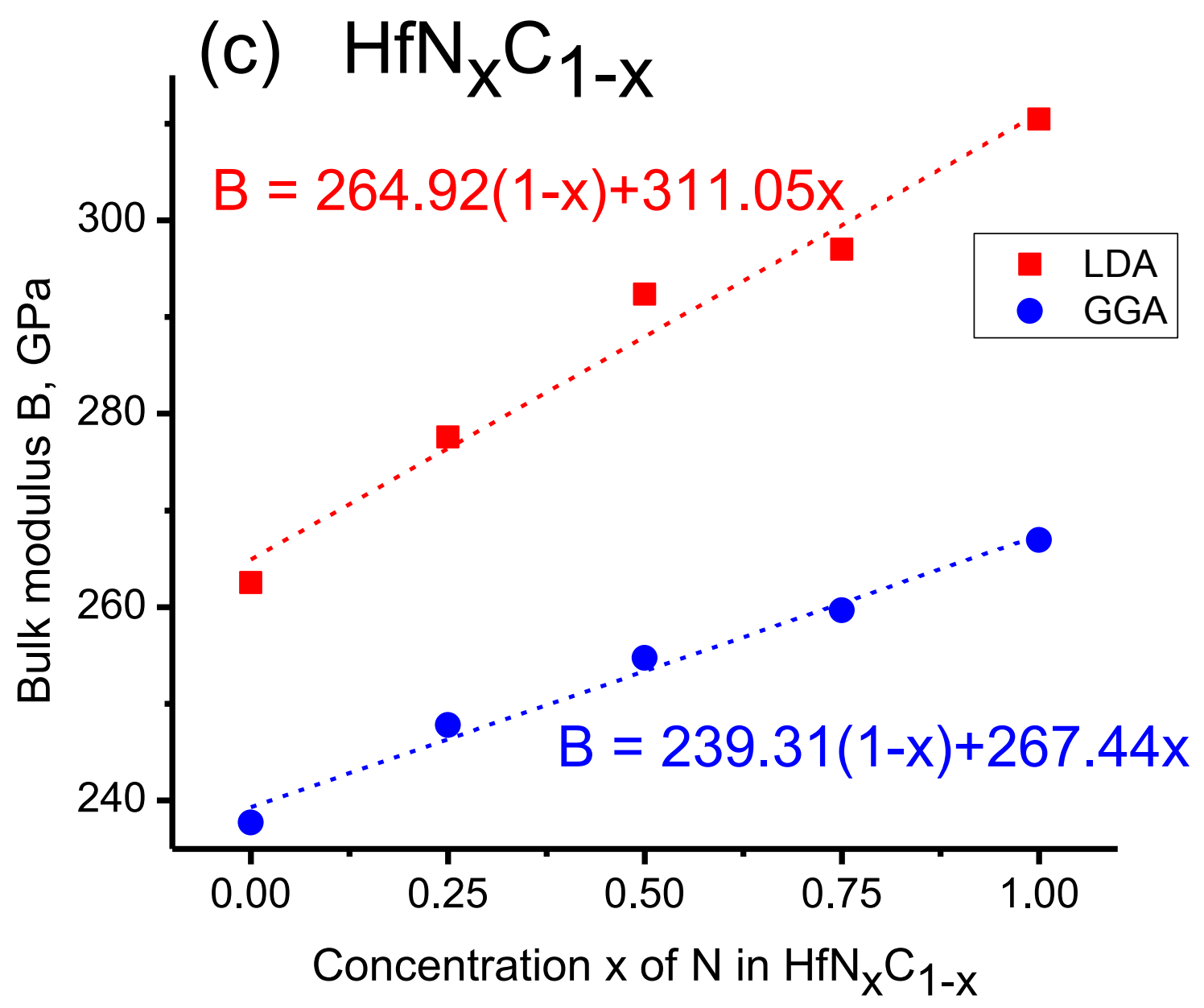

Fig. 4. Dependence of the calculated bulk modulus on the nitrogen composition *x* for $TiN_xC_{1-x}$ (a), $ZrN_xC_{1-x}$ (b) and $HfN_xC_{1-x}$ (c) carbonitrides.

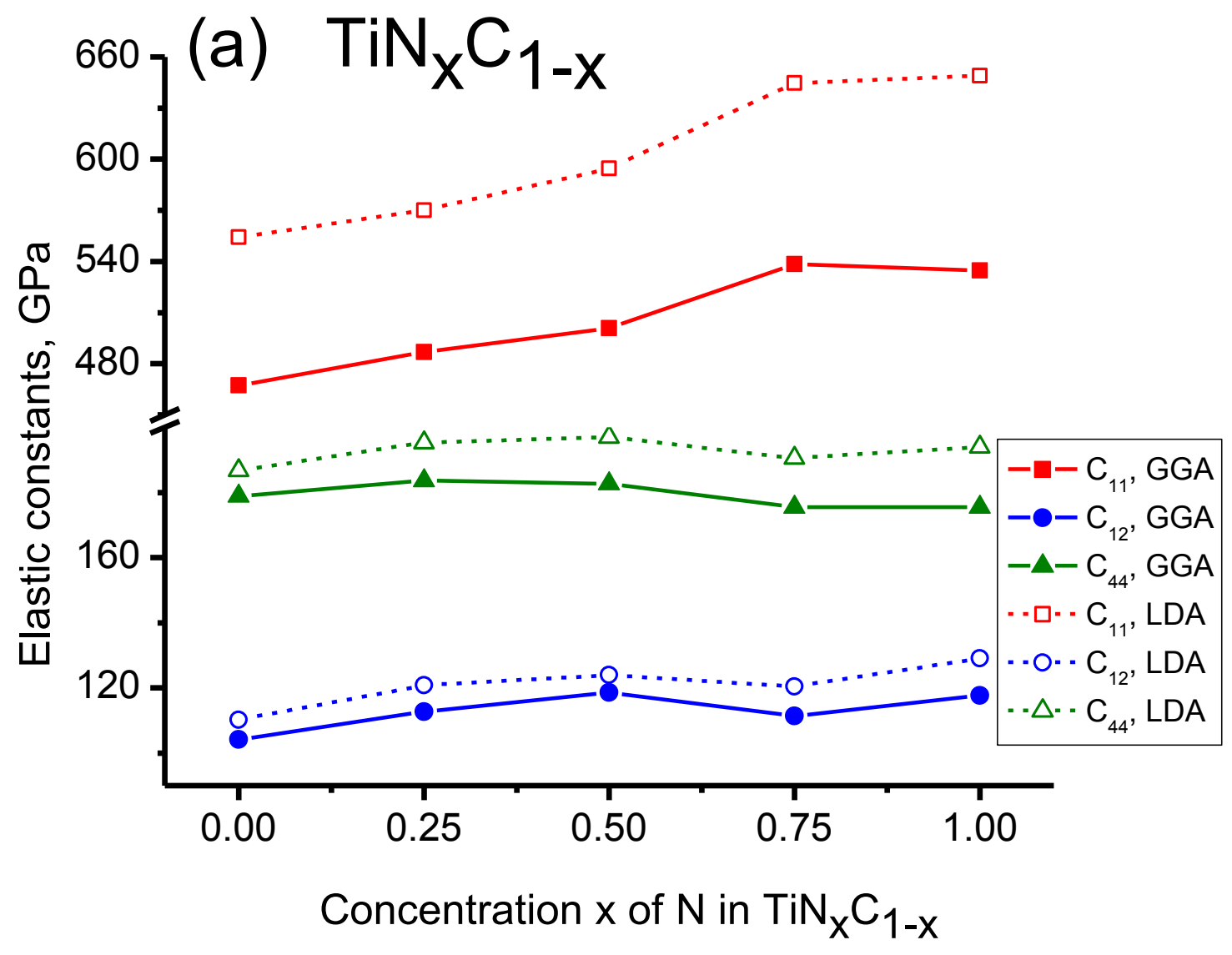

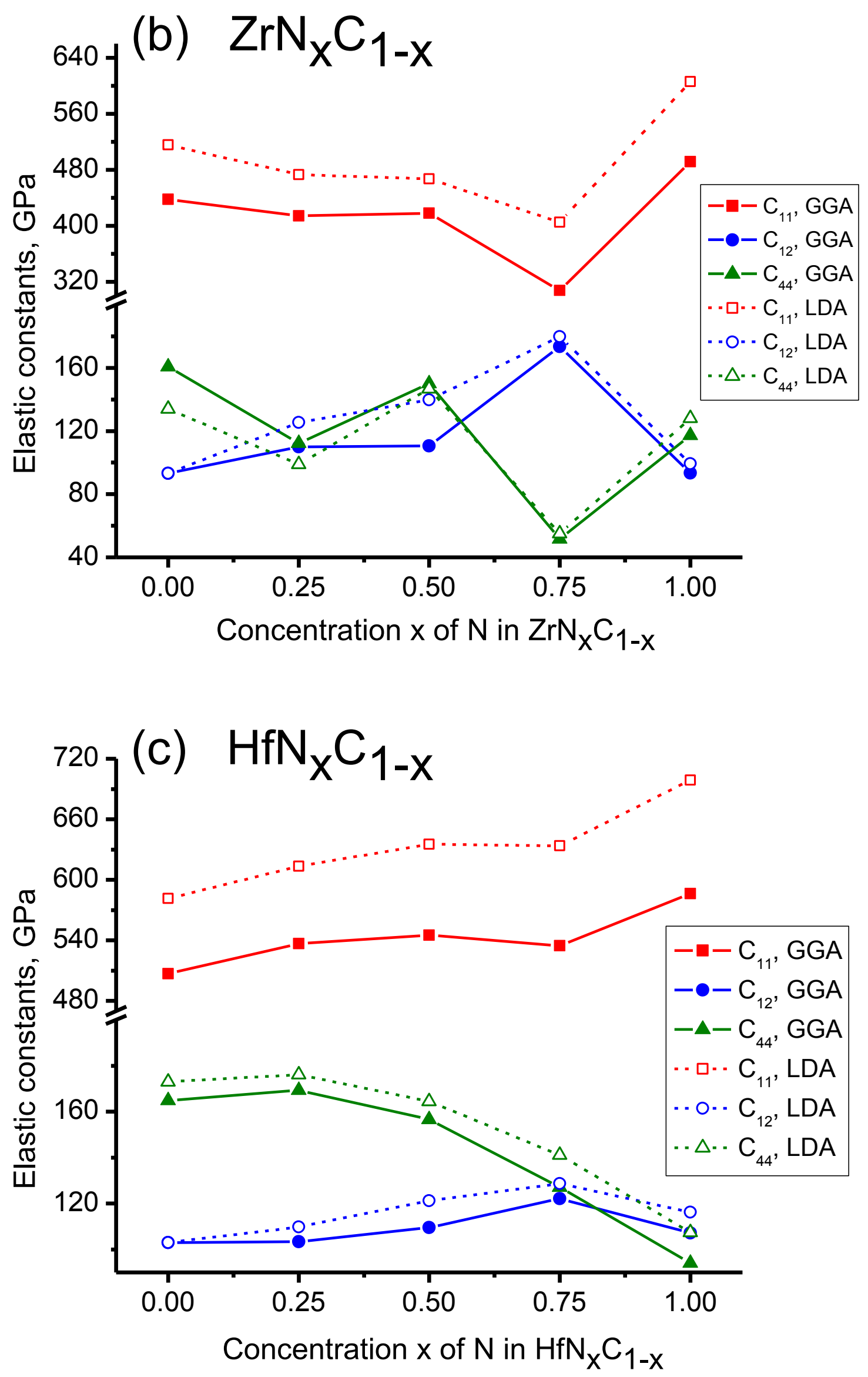

Fig. 5. Dependence of the calculated elastic constants $C_{11}$, $C_{12}$, $C_{44}$ (in GPa) on the nitrogen composition $x$ for $TiN_xC_{1-x}$ (a), $ZrN_xC_{1-x}$ (b) and $HfN_xC_{1-x}$ (c) carbonitrides.

Table 3. Elastic constants $C_{11}$, $C_{12}$, $C_{44}$ (in GPa) for three groups of the $TiN_xC_{1-x}$, $ZrN_xC_{1-x}$ and $HfN_xC_{1-x}$ carbonitrides

| Compound | Calc. (this work) | | | | | | Other data | | |
|---|---|---|---|---|---|---|---|---|---|
| | LDA | | | GGA | | | | | |
| | $C_{11}$ | $C_{12}$ | $C_{44}$ | $C_{11}$ | $C_{12}$ | $C_{44}$ | $C_{11}$ | $C_{12}$ | $C_{44}$ |
| TiC | 554.2 | 110.2 | 186.8 | 467.4 | 104.2 | 178.9 | 513[a], 504[b], 517.9[f], 597.1[f] | 106[a], 123[b], 117.3[f], 129.1[f] | 178[a], 155[b], 190[e], 164.6[f], 174.3[f] |
| $TiC_{0.75}N_{0.25}$ | 570.0 | 120.8 | 195.3 | 486.8 | 112.7 | 183.6 | 547.1[f], 621.1[f] | 123.4[f], 130.5[f] | 253[e], 191.1[f], 195.7[f] |
| $TiC_{0.50}N_{0.50}$ | 594.7 | 124.0 | 197.0 | 500.8 | 118.5 | 182.7 | 599.9[f], 653.0[f] | 115.4[f], 128.5[f] | 292[e], 175.7[f], 201.7[f] |
| $TiC_{0.25}N_{0.75}$ | 644.6 | 120.4 | 190.6 | 538.5 | 111.4 | 175.5 | 602.9[f], 659.1[f] | 126.4[f], 135.9[f] | 281[e], 174.5[f], 188.6[f] |
| TiN | 649.0 | 129.0 | 194.0 | 534.7 | 117.7 | 175.4 | 625[c], 611.7[f], 689.1[f] | 165[c], 124.1[f], 137.0[f] | 163[c], 215[e], 162.9[f], 169.0[f] |
| ZrC | 502.60 | 99.12 | 171.44 | 437.78 | 93.24 | 160.81 | 452 [g] | 102 [g] | 154 [g] |
| $ZrC_{0.75}N_{0.25}$ | 473.05 | 125.42 | 98.93 | 414.37 | 109.96 | 112.29 | 472 [g] | 104 [g] | 165 [g] |
| $ZrC_{0.50}N_{0.50}$ | 467.16 | 139.66 | 146.82 | 418.07 | 110.62 | 150.12 | 487 [g] | 106 [g] | 158 [g] |
| $ZrC_{0.25}N_{0.75}$ | 405.43 | 179.81 | 55.10 | 307.56 | 173.50 | 51.66 | 507 [g] | 106 [g] | 142 [g] |
| ZrN | 605.99 | 99.25 | 128.06 | 491.55 | 93.44 | 117.32 | 523 [g] | 107 [g] | 121 [g] |
| HfC | 581.7 | 102.9 | 173.0 | 507.1 | 103 | 164.8 | 500[a] | 105[d] | 180[a], 153[e] |
| $HfC_{0.75}N_{0.25}$ | 613.4 | 109.8 | 176.1 | 536.7 | 103.4 | 169.3 | | | 176[e] |
| $HfC_{0.50}N_{0.50}$ | 635.3 | 121.2 | 164.4 | 545.1 | 109.6 | 156.6 | | | 190[e] |
| $HfC_{0.25}N_{0.75}$ | 633.9 | 128.7 | 141.2 | 534.8 | 122.1 | 127.1 | | | 185[e] |
| HfN | 699.0 | 116.2 | 107.4 | 586.4 | 107.2 | 94.1 | 600[b], 664[e] | 109[b], 115[e] | 92[b], 154[e] |

[a] Exp., Ref. [31]
[b] Calc., Ref. [31]
[c] Exp., Ref. [40]
[d] Exp., Ref. [41]
[e] Calc., Ref. [27]
[f] Calc., Ref. [42]
[g] Calc., Ref. [29]

## 3.3. Electronic and bonding properties

The calculated band structures of all pure and mixed compounds studied in the present papers are typical of metals, similar to those published in Refs. [18, 19] and thus are not shown here for the sake of brevity. Instead, an emphasis is placed on variations of the electron density distribution and chemical bond orders with composition of the mixed carbides. The Mulliken bond order can serve as a measure of the degree of covalency or iconicity of a particular chemical bond. A high value of the bond population indicates enhanced covalency, while a low value is a manifestation of an ionic interaction between two ions [43]. Although the *absolute* values of the bond orders depend strongly on the wave functions used for the calculations, the *relative* values obtained with the same calculating settings can be compared with each other to give a qualitative description of the chemical bonding in a studied material. From this point of view, by comparing the calculated bond orders for the *M*-C and *M*-N (*M*=Ti, Zr, and Hf) bonds. For all cases, the bond order of the *M*-C bonds is greater than that one of the *M*-N bonds; therefore, that the former are more covalent than the latter, which exhibit enhanced degree of iconicity. Such a conclusion can be supported by the fact that nitrogen has a greater value of electronegativity (3.04) than carbon (2.55) [44].

An additional proof of difference in the *M*-C and *M*-N interactions comes from the shown in Fig. 6 cross-sections of electron density difference for TiC, $TiC_{0.5}N_{0.5}$, and TiN.

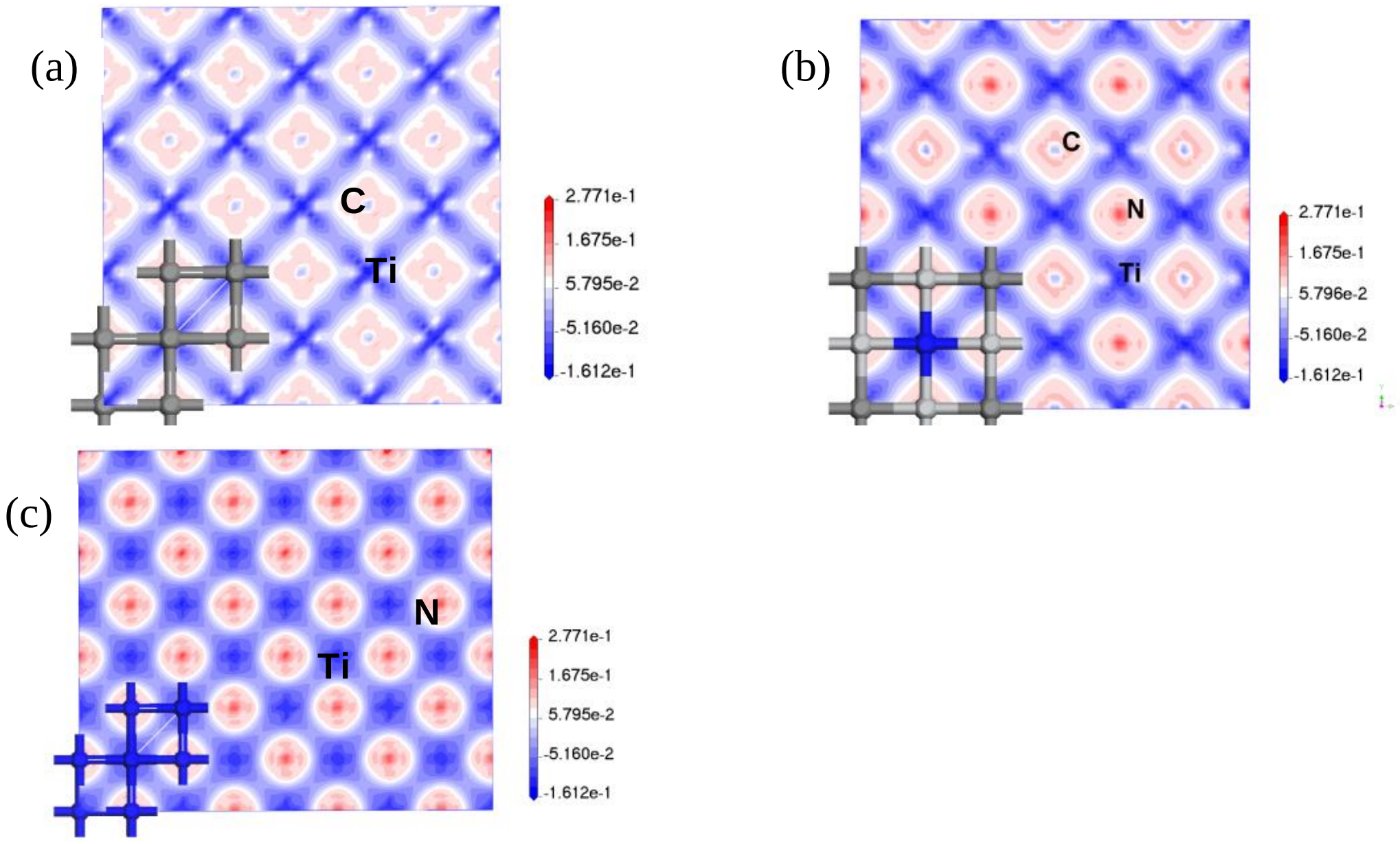

Fig. 6. Cross-sections of the electron density difference in the *ab* plane for TiC (a), $TiC_{0.5}N_{0.5}$ (b) and TiN (c).

Covalent interaction between the 3d states of Ti and 2p states of C is clearly seen in Figs. 3a and 3b through well-pronounced direction dependence of the electron density on the C ions with maxima directed toward the Ti ions. On the other hand, the electron density around the N ions tends to preserve a more spherical shape, which is a characteristic of the ionic interaction. Similar pictures for the $ZrN_xC_{1-x}$ and $HfN_xC_{1-x}$ carbonitrides are not shown for the sake of brevity.

### 3.4. Pressure effects on the elastic properties of the $TiN_xC_{1-x}$, $ZrN_xC_{1-x}$ and $HfN_xC_{1-x}$ carbonitrides

The elastic constants of all considered 15 systems were calculated in the pressure range from 0 to 50 GPa with a step of 10 GPa. Only the LDA results are shown here for the sake of brevity.

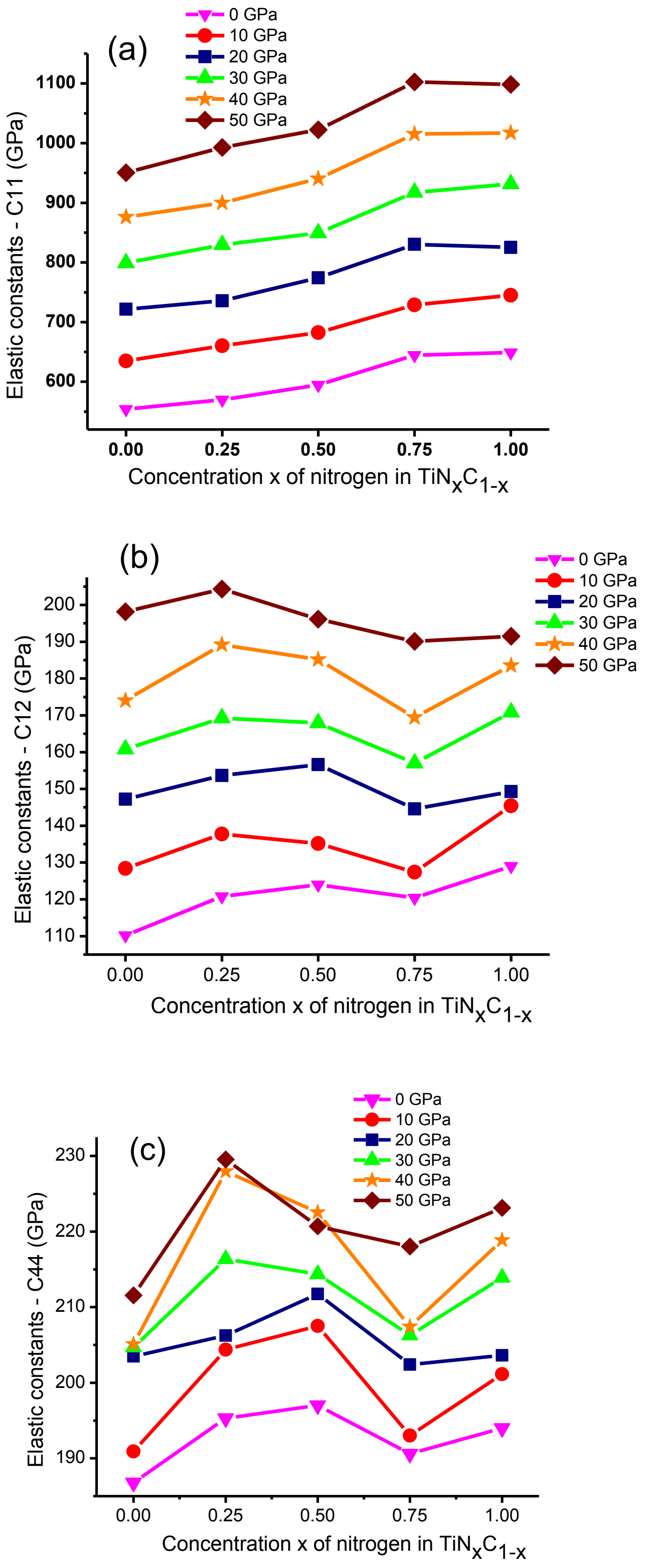

Fig. 7. Dependence of the calculated elastic constants $C_{11}$ (a), $C_{12}$ (b), and $C_{44}$ (c) (all in GPa) on the nitrogen composition $x$ for $TiN_xC_{1-x}$ in the pressure range from 0 to 50 GPa.

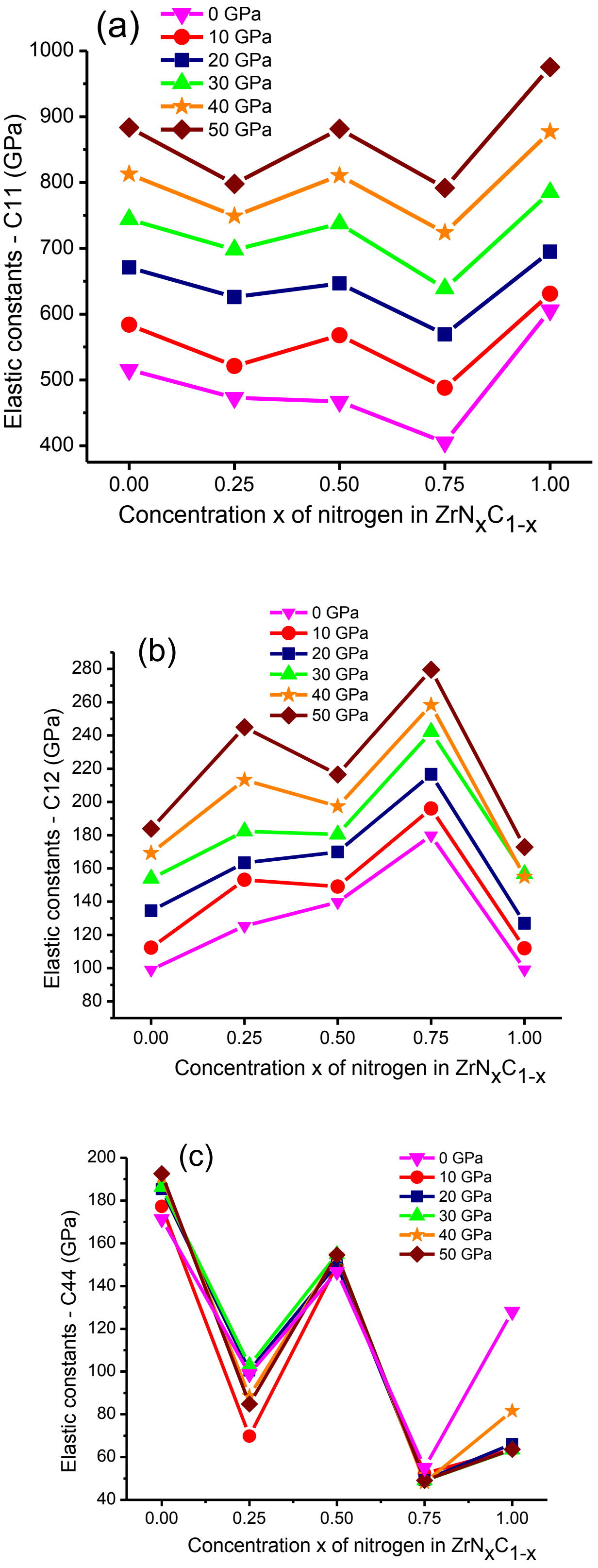

Fig. 8. Dependence of the calculated elastic constants $C_{11}$ (a), $C_{12}$ (b), and $C_{44}$ (c) (all in GPa) on the nitrogen composition $x$ for $ZrN_xC_{1-x}$ in the pressure range from 0 to 50 GPa.

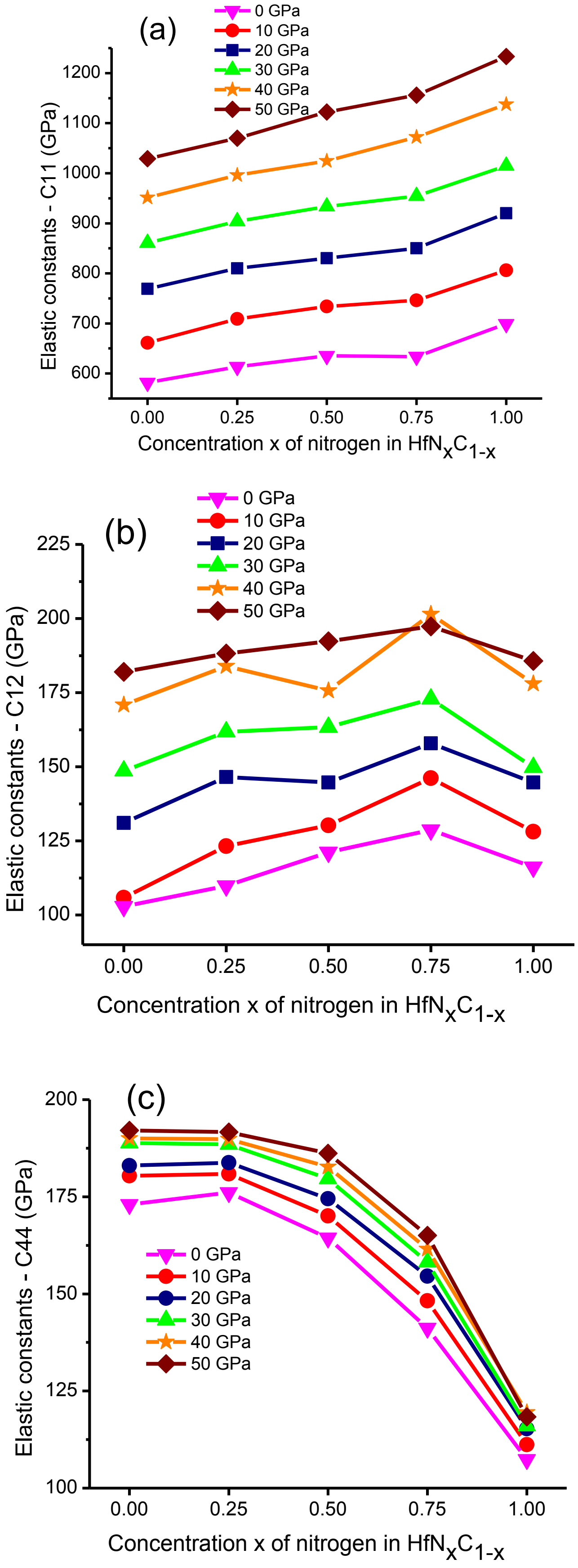

Fig. 9. Dependence of the calculated elastic constants $C_{11}$ (a), $C_{12}$ (b), and $C_{44}$ (c) (all in GPa) on the nitrogen composition $x$ for $HfN_xC_{1-x}$ in the pressure range from 0 to 50 GPa.

Figs. 7, 8, and 9 show the variation of the $C_{11}$, $C_{12}$, and $C_{44}$ elastic constants for $TiN_xC_{1-x}$, $ZrN_xC_{1-x}$, and $HfN_xC_{1-x}$ in the pressure range from 0 to 50 GPa. For each nitrogen concentration, the values of the elastic constants increase with pressure. As for the variation with the nitrogen concentration - only the $C_{11}$ constant (related to the longitudinal compression) shows monotonic increase when moving from neat carbide to neat nitride for the Ti- and Hf-based compounds. For $ZrN_xC_{1-x}$ the value of $C_{11}$ constant for ZrN is greater than for ZrC, although the variation is not monotonic and shows a minimum for $x$=0.75, which is repeated for all considered pressures. It can be also noticed in Fig. 8 that for the $ZrN_xC_{1-x}$ carbonitride increase of the $C_{11}$ constant with the nitrogen concentration is followed by a decrease of the $C_{12}$ constant (related to the transversal expansion appearing as a result of the longitudinal compression), and vice versa. The shear constant $C_{44}$ does not exhibit any systematic behavior in the case of $TiN_xC_{1-x}$ and $ZrN_xC_{1-x}$; however, it decreases with increasing the nitrogen concentration in $HfN_xC_{1-x}$.

In spite of such complicated behavior of the individual components of the elastic tensor for the considered carbonitrides, the bulk moduli $B$, which can be calculated as $B = (C_{11} + 2C_{12})/3$, all exhibit a linear increase with nitrogen concentration at all pressures (Fig. 10).

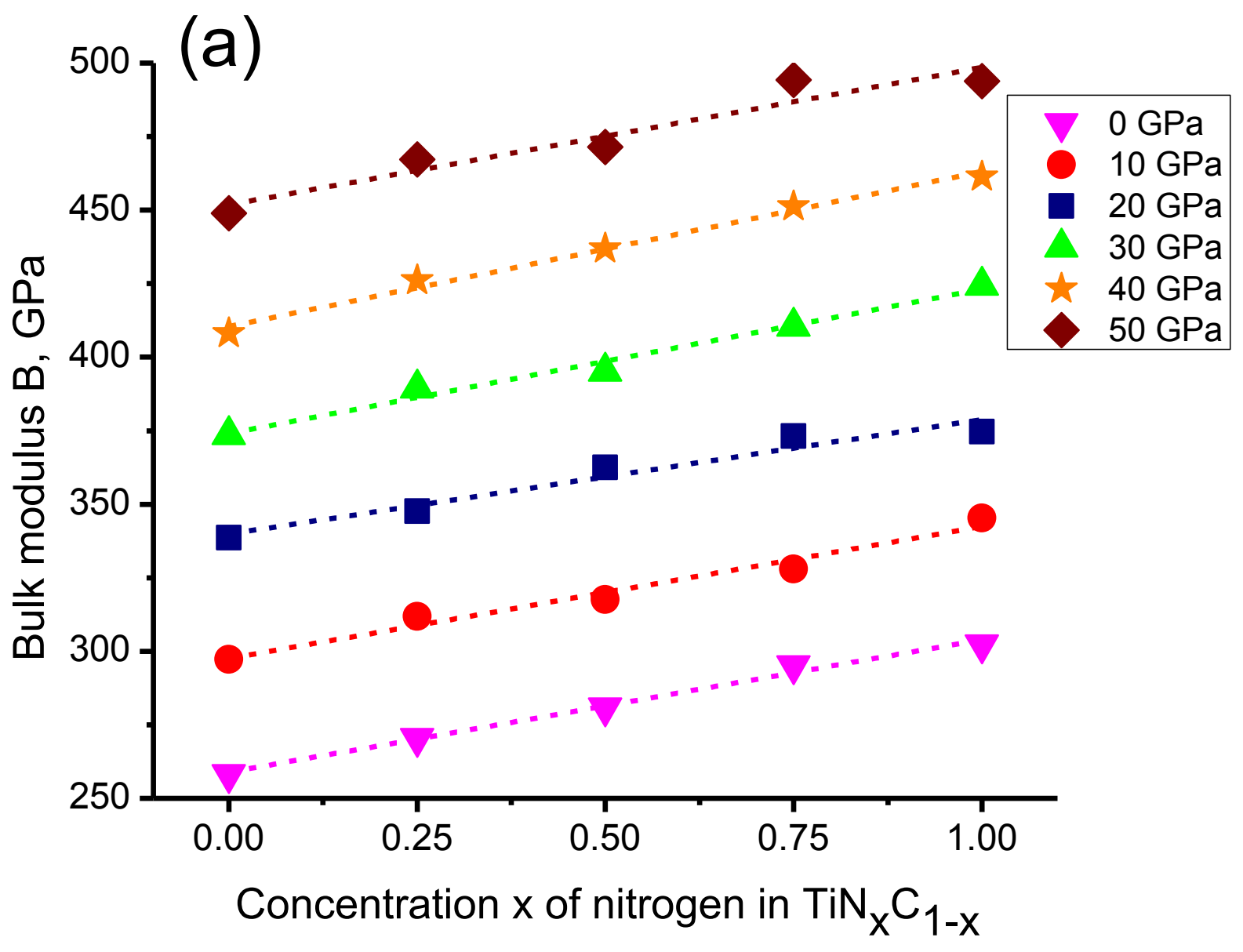

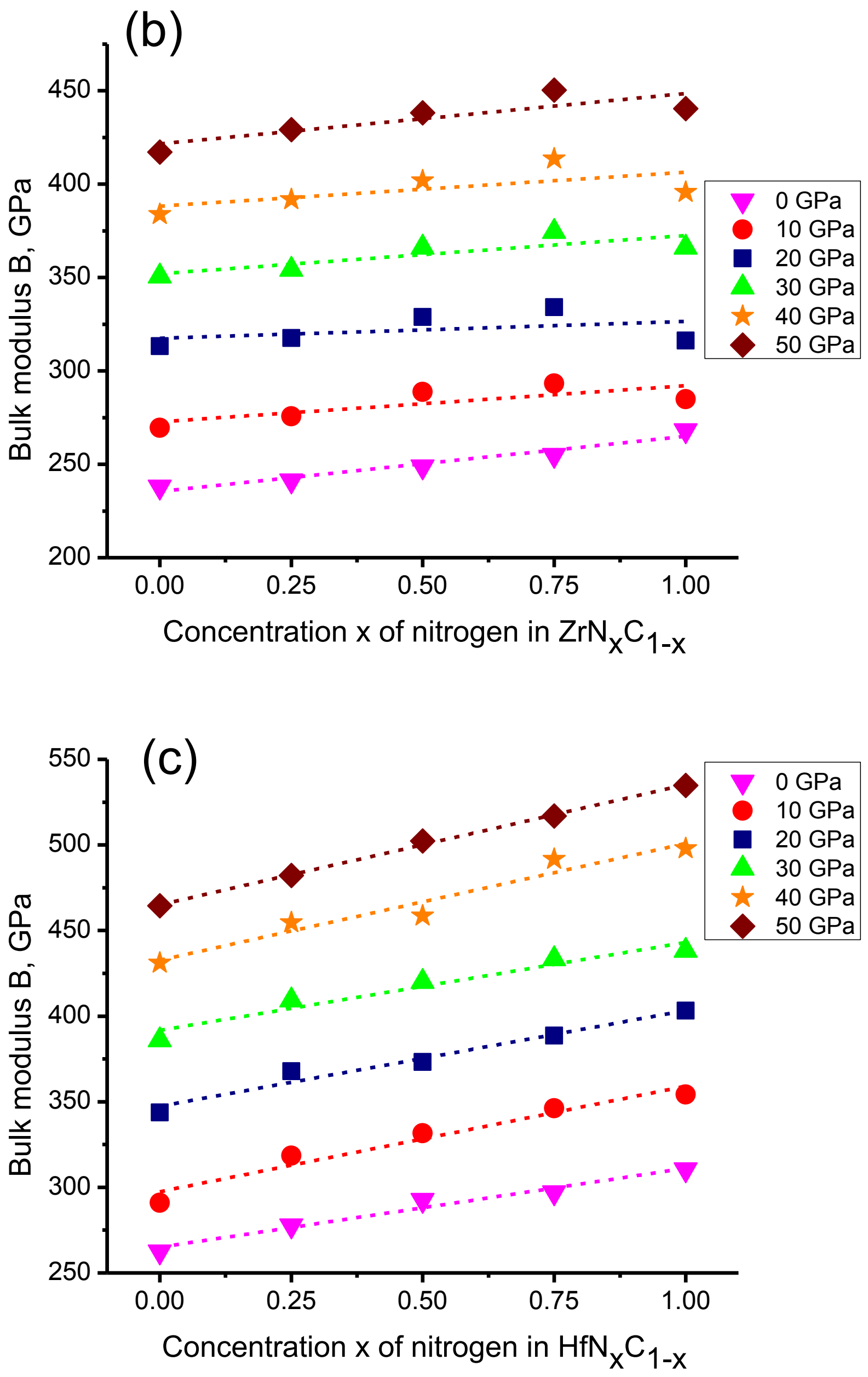

Fig. 10. Dependence of the calculated bulk modulus *B* (in GPa) on the nitrogen composition *x* for $TiN_xC_{1-x}$ (a), $ZrN_xC_{1-x}$ (b) and $HfN_xC_{1-x}$ (c) in the pressure range from 0 to 50 GPa with linear fits (their equations are given in Table 4).

Table 4. Bulk moduli *B* (in GPa) for three groups of the $TiN_xC_{1-x}$, $ZrN_xC_{1-x}$ and $HfN_xC_{1-x}$ carbonitrides as linear functions of the nitrogen concentration *x* (equations of straight lines from Fig. 10).

| Pressure, GPa | Calculated (this work) LDA |
|---|---|
| | $TiN_xC_{1-x}$ |
| 0 | 258.85 (1-*x*)+304.0 *x* |
| 10 | 297.53 (1-*x*)+342.46 *x* |
| 20 | 339.84 (1-*x*)+378.79 *x* |
| 30 | 374.02 (1-*x*)+423.16 *x* |
| 40 | 410.40 (1-*x*)+463.14 *x* |
| 50 | 451.76 (1-*x*)+ 498.51 *x* |
| | $ZrN_xC_{1-x}$ |
| 0 | 233.03 (1-*x*)+265.71 *x* |
| 10 | 272.76 (1-*x*)+ 292.11 *x* |
| 20 | 317.49 (1-*x*)+ 326.48 *x* |
| 30 | 352.03 (1-*x*)+ 372.55 *x* |
| 40 | 388.23 (1-*x*)+ 406.40 *x* |
| 50 | 421.54 (1-*x*)+ 448.57 *x* |
| | $HfN_xC_{1-x}$ |
| 0 | 264.92 (1-*x*)+311.05 *x* |
| 10 | 297.43 (1-*x*)+359.08 *x* |
| 20 | 347.35 (1-*x*)+403.29 *x* |
| 30 | 391.68 (1-*x*)+443.06 *x* |
| 40 | 432.62 (1-*x*)+500.90 *x* |
| 50 | 465.01 (1-*x*)+535.24 *x* |

Table 4 collects all equations of linear fits from Fig. 10, which give an opportunity to calculate the value of the bulk modulus for the studied carbonitrides with any concentration of nitrogen and at varying hydrostatic pressure. The pressure derivatives of the calculated bulk moduli are all close to 4 - a typical value for solids. Fig. 11 and Table 5 show the variation of the bulk moduli with pressure and equations of linear fits, respectively.

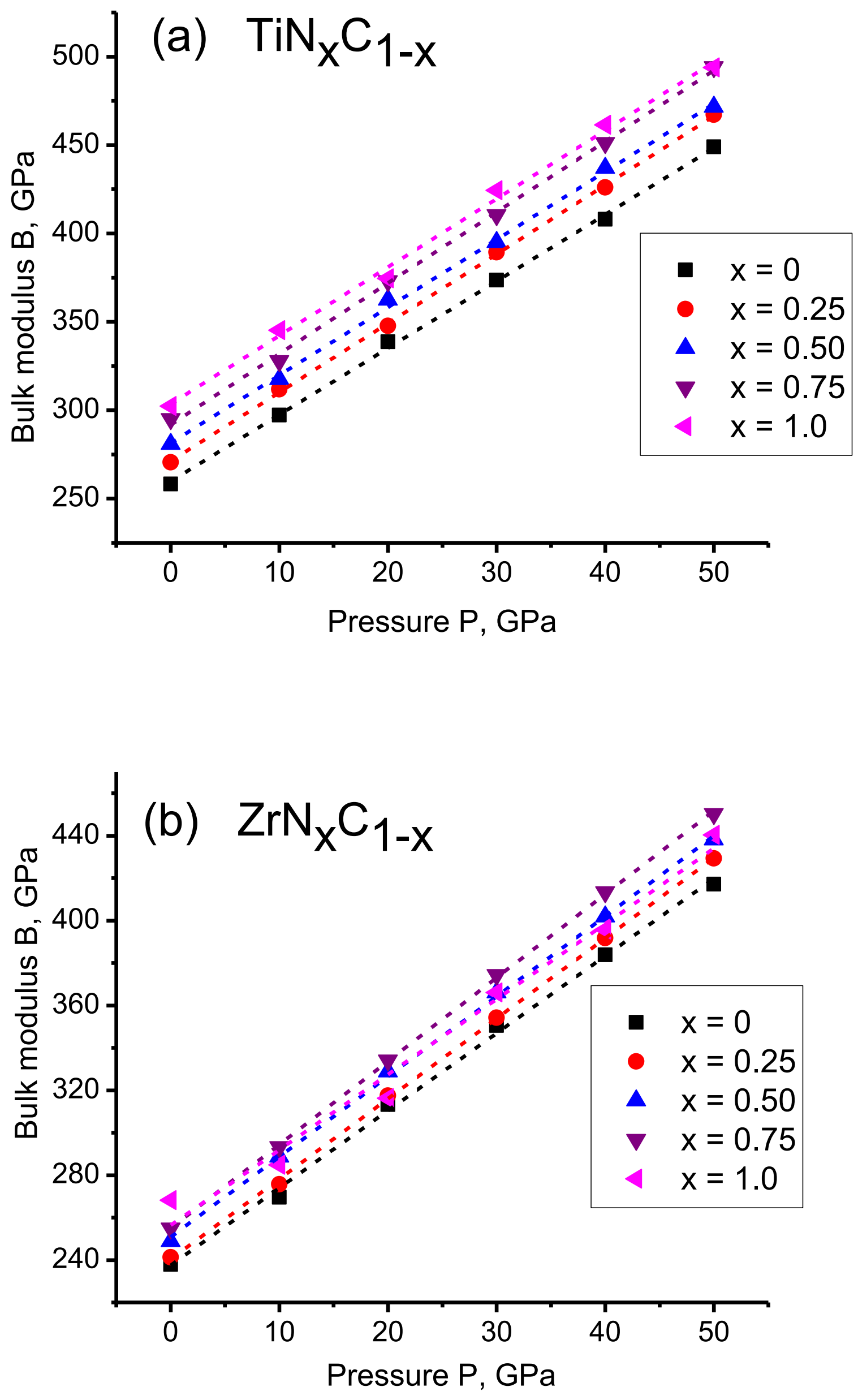

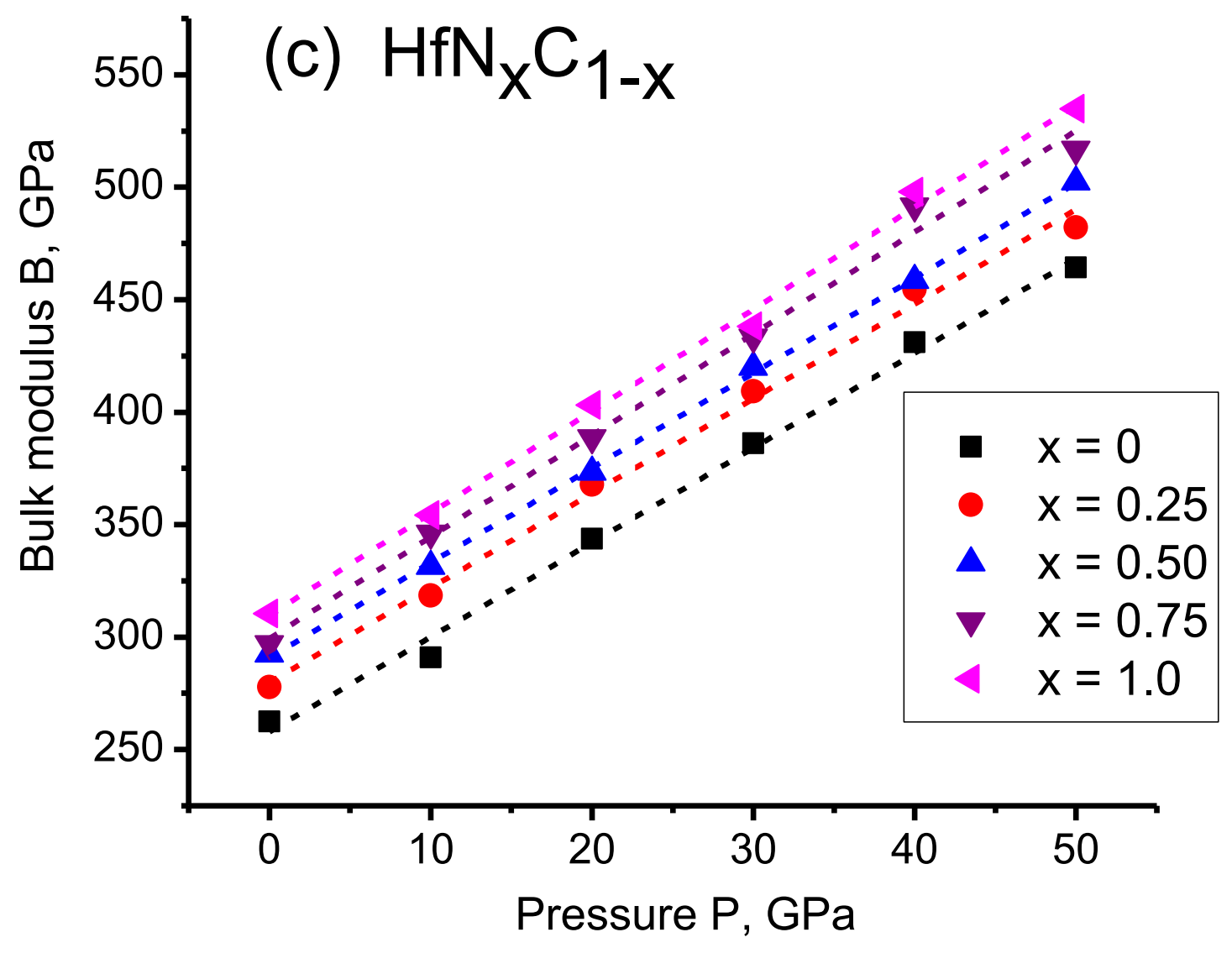

Fig. 11. Dependence of the calculated bulk modulus *B* (in GPa) on the external pressure *P* for $TiN_xC_{1-x}$ (a), $ZrN_xC_{1-x}$ (b) and $HfN_xC_{1-x}$ (c) in the pressure range from 0 to 50 GPa with linear fits (their equations are given in Table 5).

Table 5. Bulk moduli *B* (in GPa) for as functions of pressure *P* for three groups of the $TiN_xC_{1-x}$, $ZrN_xC_{1-x}$ and $HfN_xC_{1-x}$ carbonitrides as linear functions of the nitrogen concentration *x* (equations of straight lines from Fig. 11).

| Nitrogen concentration *x* | Calculated (this work) LDA |
|---|---|
| | $TiN_xC_{1-x}$ |
| 0 | 259.73+3.77 *P* |
| 0.25 | 271.11+3.91 *P* |
| 0.50 | 281.44+3.84 *P* |
| 0.75 | 291.84+4.01 *P* |
| 1.00 | 303.50+3.87 *P* |
| | $ZrN_xC_{1-x}$ |
| 0 | 237.54+3.65 *P* |
| 0.25 | 240.35+3.78 *P* |
| 0.50 | 250.91+3.78 *P* |
| 0.75 | 255.07+3.94 *P* |
| 1.00 | 256.42+3.55 *P* |
| | $HfN_xC_{1-x}$ |
| 0 | 257.97+4.21 *P* |
| 0.25 | 279.80+4.21 *P* |
| 0.50 | 290.90+4.22 *P* |
| 0.75 | 299.44+4.51 *P* |
| 1.00 | 309.70+4.54 *P* |

## 4. Conclusions

First-principles calculations of the structural, electronic and elastic properties of three carbonitrides $TiN_xC_{1-x}$, $ZrN_xC_{1-x}$, $HfN_xC_{1-x}$ ($0 \leq x \leq 1$) performed in this paper allowed to follow how the lattice constant, density, bulk modulus, elastic tensor components depend on the nitrogen concentration. It was shown that the Vegard's law holds true for all above-mentioned quantities, except for the elastic constants $C_{11}$, $C_{12}$, and $C_{44}$. The linear approximations of the lattice constant, density and bulk modulus can be used for reliable estimation of their values for any value of nitrogen concentration $x$ from 0 to 1.

Consideration of the bonding properties of the studied materials, based on the calculated distribution of the electron density difference in the space between atoms in crystal lattices, led to the conclusion that gradual substitution of carbon by nitrogen enhances ionic character of chemical bonds on account of decreased covalency.

Pressure effects on the elastic properties were modeled by optimizing the crystal structure at elevated hydrostatic pressure with subsequent calculations of the elastic constants. In this way, dependence of the bulk moduli on the nitrogen concentration was obtained for different values of pressure: 0, 10, 20, 30, 40 and 50 GPa, thus giving a complete description of the bulk moduli variation for the considered carboitrides as functions both of pressure and nitrogen concentration.

The data obtained in the present paper can be used for analysis of behavior of the selected three carbonitrides at various conditions of their applications.

## Acknowledgment

This study was supported by the European Union through the European Regional Development Fund (Centre of Excellence "Mesosystems: Theory and Applications", TK114) and European Social Fund's Doctoral Studies and Internationalization Programme DoRa. We also thank Dr. G.A. Kumar (University of Texas at San Antonio) for allowing to use the Materials Studio package. We are grateful to the High Performance Computing Centre, University of Tartu, for computational facilities.